\documentclass[superscriptaddress,reprint,amssymb, nobibnotes, aps, pra]{revtex4-2}

\DeclareMathAlphabet{\mathpzc}{OT1}{pzc}{m}{it}

\usepackage{lipsum}
\usepackage{graphicx}
\usepackage[caption=false]{subfig}
\usepackage{epsfig}
\usepackage{epstopdf}
\usepackage{amsmath}
\usepackage{amsfonts}
\usepackage{ upgreek }
\usepackage{float}
\usepackage{enumerate}
\usepackage{array}
\usepackage{pifont}
\usepackage{comment}
\usepackage[11pt]{moresize}
\usepackage{datetime}
\usepackage{multirow}

\usepackage{mathrsfs}
\newcolumntype{C}[1]{>{\centering\let\newline\\\arraybackslash\hspace{0pt}}m{#1}}
\newcolumntype{N}{@{}m{0pt}@{}}

\begin{document}



\title{Photonic-Spin governed absorption in a mu-near-zero subwavelength gyromagnetic cylinder}



\author{Rajarshi Sen}
\affiliation{Department of Electronics and Electrical Communication Engineering, Indian Institute of Technology Kharagpur, Kharagpur, West Bengal, India}
\author{Sarang Pendharker}
\affiliation{Department of Electronics and Electrical Communication Engineering, Indian Institute of Technology Kharagpur, Kharagpur, West Bengal, India}
\email[]{sarang@ece.iitkgp.ac.in}



\date{\today}

\begin{abstract}

Realizing non-reciprocal scattering and absorption from subwavelength elements has several applications across the electromagnetic frequency spectrum. Here we show that gyromagnetic cylindrical rods with subwavelength dimensions exhibit strong absorption with diminished scattering in a narrow frequency band for an incident plane wave. We show that the strong narrowband absorption depends on the photonic-spin of the incident wave. Further, it is shown that this spin-dependent absorption leads to non-reciprocal reflection and absorption characteristics of an array of gyromagnetic cylinders suspended over a reflecting plane. The theoretical claims of strong spin-dependent absorption in subwavelength elements is validated by experimentally demonstrating nonreciprocal absorption and transmission from a single gyromagnetic rod placed in a substrate integrated waveguide. 

\end{abstract}

\maketitle

\section{Introduction}

Developing subwavelength components for the control and manipulation of electromagnetic waves is essential for realizing compact and integrated photonic and microwave devices. Investigation of scattering and absorption from structures with subwavelength features has been attracting considerable interest, primarily for this reason \cite{Kivshar2022}. While the fundamental theory of scattering and absorption has been known for long \cite{Huffman_Bohren_scattering_book}, our increasing capability to engineer the material properties, coupled with the enhanced precision over structural and geometric features, is driving the research on subwavelength components over the past couple of decades. Several papers have reported the enhanced \cite{Mahmoud2014} and tunable \cite{Mahmoud_2023,Hou_2023} scattering in engineered epsilon-near-zero (ENZ) structures \cite{Issah_2023}. More recently Mie-resonance-based control of scattering properties has been reported with subwavelength high dielectric constant materials \cite{Babicheva:24}. The primary goal of these proposed structures is to achieve control over the frequency and direction of scattering. In \cite{Mie_anisotropic_scattering}, Mie scattering from nanoparticles close to a substrate has been used to estimate the optical anisotropy of the substrate. Recently, in \cite{SKV_2024_paper_perfect_mirror}, almost perfect reflection with engineered subwavelength features has been proposed. In \cite{Loran_2023}, a transfer matrix approach to analytically analyze scattering from an ensemble of anisotropic nonmagnetic point scatterers is presented. 

Scattering of electromagnetic beams from engineered and natural particles has also been an active area of research \cite{Li_23}. The directional nature of scattering from chiral particles and the resultant directional optical forces have been reported in \cite{Wang2014}. 
\cite{Kalhor2016} shows that this traverse optical force is linked to spin-momentum locking \cite{VanMechelen2016} in EM waves, and \cite{Sarang_2018} explores its consequences in non-reciprocal and moving waveguide structures. In \cite{transverse_spin_mie_scatter,Cao2024} the presence of transverse spin in the vicinity of Mie scatterer has been investigated. In \cite{Zurita2021} it is shown that the direction of scattering from Mie scatterer can depend on the handedness of a circular dipole in its vicinity. Apart from chiral particles, asymmetric scattering is observed from gyrotropic scatterers as well. In \cite{Valagiannopoulos_zeeman_scatter}, the asymmetric scattering from magnetically biased gyrotropic cylinder has been investigated. Non-reciprocal scattering properties of an ensemble of gyroelectric scatterers have been investigated in \cite{Arruda2016,Sadrara2024}. In \cite{subwavelength_antenna_directive_scattering}, the highly directive nature of radiating from a current element close to a layered gyrotropic cylinder is presented, and it is shown that the directivity and the direction of radiation can be controlled by engineering the layered gyrotropic structure.

Apart from the scattering, control over the absorption characteristic of particles and engineered structures \cite{Radi2015,PhysRevA.94.063841,Kotlicki2014,Baranov2017} is also of importance in several applications, ranging from efficient trapping of EM energy for solar cell applications \cite{C3TA13655H}, to heating, anti-reflection sheets and stealth applications. Absorption is of interest at microwave frequencies, primarily for shielding \cite{Lv2023} and stealth \cite{Dhumal2023,Payne2021, Feng2024} applications. Recently composites with coinciding ENZ and mu-near-zero (MNZ) frequencies in the RF-microwave range have been reported \cite{Dai2024}. In \cite{gyro_plasma_scatter}, directional scattering of microwave by a magnetically biased column of plasma was experimentally demonstrated. Despite such wide investigations, the role of photonic-spin of the incident wave on the absorption and scattering properties of gyrotropic particles has not been explored.

In this paper we report photonic-spin-dependent absorption and scattering in subwavelength gyromagnetic scatters and experimentally demonstrate it by realizing a substrate integrated isolator using a single subwavelength gyromagnetic absorber. We show that depending on the spin of the incident wave, a gyromagnetic absorber can exhibit almost perfect absorption at Mu-near-zero (MNZ) frequency. A rigorous investigation of the spin dependence of this absorption is presented. It is shown that a desired spin in the incident wave can be generated by superposition of two out-of-phase waves incident at oblique angles. By controlling the phase difference between these waves, the photonic spin of the magnetic field can be controlled. Further, it is shown that a more practical way of generating and controlling the spin of wave incident on the scatterer is using simple reflection from a reflector and controlling the angle of incidence. We show that placing subwavelength gyromagnetic rods near a reflector results strong direction-dependent absorption as a consequence of spin-dependent absorption. We validate our claims by experimentally demonstrating spin- and direction-dependent absorption in  subwavelength magnetically biased ferrite rod embedded inside a substrate integrated waveguide. 

The paper is organized as follows. Section~II presents the scattering and absorption properties of subwavelength gyromagnetic cylinder near MNZ frequency. Section~III shows that the absorption efficiency depends on the photonic-spin of the incident wave. In Section~IV, we show how this spin-dependent absorption leads to non-reciprocal reflection. Finally in Section~V, the claims on spin-dependent absorption are experimentally validated by realizing an absorbing MNZ subwavelength isolator.

\section{Plane wave absorption and scattering by MNZ gyromagnetic cylinder}

\begin{figure}
    \centering
    \includegraphics[width=\columnwidth]{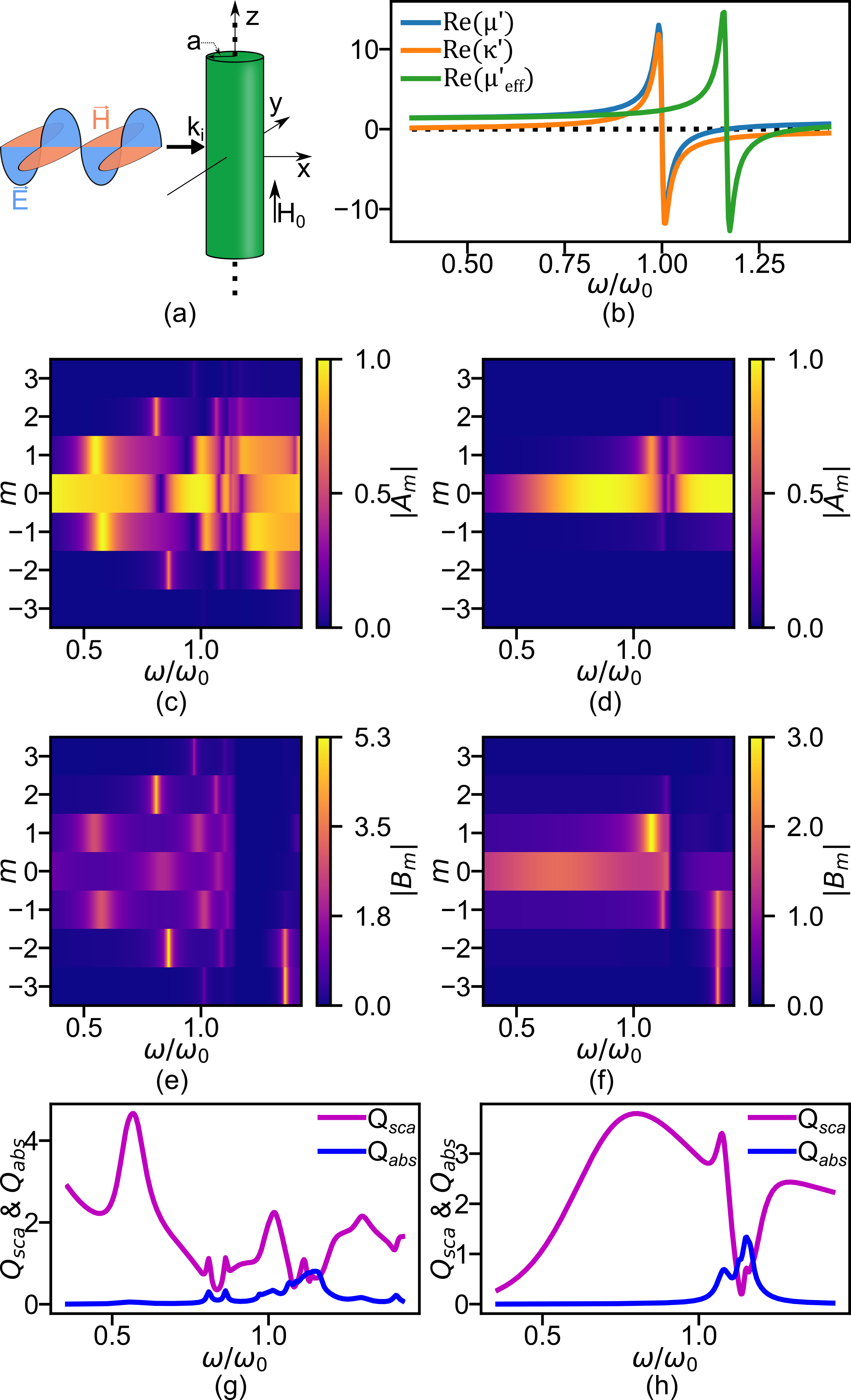}
    \caption{(a) Schematic of TE wave incident over the $\hat{z}$-biased ferrite rod along $\hat{x}$-axis, (b) Real comments of medium parameters $\mu^\prime$, $\kappa^\prime$ and $\mu_{eff}$ vs. frequency. $\mu_{eff}$ makes first zero crossing at $\omega=1.17\omega_0$. (c) and (d) $|A_m|$ variation with respect to $f$ and $m$ for $a=0.15\lambda_0$ and $a=0.05\lambda_0$, respectively. (e) and (f) $|B_m|$ variation with respect to $f$ and $m$ for $a=0.15\lambda_0$ and $a=0.05\lambda_0$, respectively. Scattering $Q_{sct}$ and Absorption efficiencies $Q_{abs}$ versus frequency, corresponding to (g) $a=0.15\lambda_0$ and (h) $a=0.05\lambda_0$, respectively.}
    \label{fig:fig1}
\end{figure}

Scattering of a uniform plane wave from a right circular cylinder has been extensively investigated, and the underlying formulation is given in the book \cite{Huffman_Bohren_scattering_book}. Computation of scattering from a gyromagnetic cylinder also follows a similar procedure, except that for the fields profile and resonance inside the gyromagnetic cylinder, we need to consider the isofrequency surfaces of the gyromagnetic medium. Here, we consider a gyromagnetic medium biased along $\hat{z}$ axis with permeability tensor in the Cartesian form given by, 
\begin{equation}
    \overset{\leftrightarrow}{\mu}_r=
    \begin{bmatrix}
        \mu^\prime&-j\kappa^\prime&0\\
        j\kappa^\prime&\mu^\prime&0\\
        0&0&1\\
    \end{bmatrix}
\end{equation}
\begin{table}
    \centering
    \caption{Ferrite material parameters.}
    \begin{tabular}{cc}
    \hline
    \hline
        Parameter & Value\\
    \hline
        Saturation Magnetization ($4\pi M_s$) & $1800$ Gauss\\
        Land\'e g factor ($g$) & 2 \\
        Dielectric permittivity ($\epsilon_r$) & 15\\
        Linewidth ($\Delta H$) & $75$ Oe \\
        Loss tangent ($\tan\delta$) & 0\\
    \hline
    \hline
    \end{tabular}
    \label{tab:default_aprms}
\end{table}
Here $\mu^\prime$ is the relative permeability along the $\hat{x}$ and $\hat{y}$ directions, and $\kappa^\prime$ is the gyrotropic term. The equations for $\mu^\prime$ and $\kappa^\prime$ are
\begin{equation}
    \begin{split}
        \mu^\prime&=1+\frac{\omega_0\omega_m}{\omega_0^2-\omega^2}\\
        \kappa^\prime&=\frac{\omega\omega_m}{\omega_0^2-\omega^2}\\
    \end{split}
\end{equation}
Here $\omega_0=\mu_0\gamma H_0$ and $\omega_m=\mu_0\gamma M_s$. $\gamma=(gq)/(2m_e)$ is the gyromagnetic ratio, $g$ is the land\'e g factor, $q$ is the charge of an electron, and $m_e$ is the mass of an electron. $H_0$ the applied bias field strength, and $M_s$ is the saturation magnetization of the ferrite. The sign of $\kappa^\prime$ can be reversed with the reversal of the direction of magnetic bias. Losses are incorporated in the formulation of $\mu^\prime$ and $\kappa^\prime$ by substituting $\omega_0$ with $\omega_0-j\alpha\omega$, where the damping factor $\alpha$ is equal to $(\mu_0\gamma\Delta H)/(2\omega)$. The parameter $\Delta H$ is the linewidth of the ferrite signifying the gyromagnetic loss. The other source of loss is the dielectric loss tangent $\tan\delta$, and the resulting complex dielectric permittivity is easy to calculate. For the material properties and magnetic bias value mentioned in Table~\ref{tab:default_aprms} and magnetic bias of $\vec{H_0}=5000\hat{z}$ Oe, the values of $\mu^\prime$, $\kappa^\prime$ and $\mu_{eff} = (\mu^{\prime2}-\kappa^{\prime2})/\mu^\prime$ are shown in Fig~\ref{fig:fig1}(b). It can be easily shown that the $\overset{\leftrightarrow}{\mu}_r$ tensor has similar from in the cylindrical coordinates as well.


This gyromagnetic material can support two modes of propagation in the $x-y$ plane. The propagation properties  of these modes are governed by the corresponding isofrequency curves (IFCs). In \cite{Sen_2022}, we had shown that the two IFCs exhibit different photonic-spin characteristics and how $\kappa^\prime$ effects the existence of these IFCs. Here we consider the incident uniform plane wave with electric field along the $\hat{z}$ direction and the magnetic field to be along the $x-y$ plane, as shown in Fig.~\ref{fig:fig1}(a).  For this mode, the propagation constant is given by $k=k_0\sqrt{\mu_{eff}\epsilon_r}$. We consider this mode because it has a transverse magnetic field spin in the $\hat{z}$ direction, with magnetic spin defined by $\vec{S}_h = \textrm{Im}(\vec{H}^*\times\vec{H})$. Inside the cylinder, the $\hat{r}$ and $\hat{\phi}$ components of magnetic field are not independent and are coupled by the gyrotropic term $\kappa^\prime$. Therefore, when a uniform plane wave is incident on such a cylinder, owing to the the gyrotropy and the boundary conditions, the scattered field will also have both $\hat{r}$ and $\hat{\phi}$ components. As a result, the scattered wave will also have out-of-phase $H_x$ and $H_y$ components resulting in photonic spin in the magnetic field in the z-direction $\vec{S}_h=S_z\hat{z}$. Therefore, a gyrotropic cylinder causes asymmetric scattering as well as introduces a spin in the scattered field. 

Figure~\ref{fig:fig1}(c) and (d) show the scattering coefficients of scattered field for rod radius of $a=0.15 \lambda_0$ and $a=0.05\lambda_0$, respectively. $\lambda_0$ is the free-space wavelength corresponding to frequency $\omega_0$. The corresponding field coefficients inside the rod are shown in Fig.~\ref{fig:fig1}(e) and (f). $m$ is the index of Bessel and Hankel functions. The required equations and formulation are given in Appendix~A. The values of $A_m$ are responsible for the scattering cross-section $C_{sct}$ and scattering efficiency $Q_{sct}$. Whereas the $B_m$s are responsible for absorption cross-section $C_{abs}$ and absorption efficiency $Q_{abs}$. The expressions and details of computation are given in Appendix~A. It can be seen that for higher rod radius, the structural resonance modes dominate the response. Due to gyrotropy, the frequency spectrum for positive and negative values of $m$ becomes asymmetric. This is the reason for Zeeman splitting and asymmetric scattering pattern as mentioned in \cite{Valagiannopoulos_zeeman_scatter}. Further, it can be observed for larger rod size (from Fig.~\ref{fig:fig1}(c) and (e)) that while the positive and negative $m$ resonances are offset the corresponding positive and negative modes exists within the frequency range. On the other hand for smaller rod radius of $a=0.05\lambda_0$, as seen in Fig.~\ref{fig:fig1}(d) and (f), this asymmetry is enhanced, and the frequency spectrum is a lot more skewed. We can see that there is a possibility of only positive or negative modes existing over the frequency band of interest. This is because for small rod-radius, the material resonance dominates, which due to gyrotropy, has a strong preferred sense of spin. The scattering and absorption efficiencies corresponding to rod radius of $a=0.15\lambda_0$ and $a=0.05\lambda_0$ are shown in Fig.~\ref{fig:fig1}(g) and (h) respectively. Interestingly, in Fig.~\ref{fig:fig1}(h), there is a narrow frequency band, near MNZ frequency, where the scattering efficiency drops to near zero while the absorption efficiency peaks. This is the frequency of interest to us for realizing MNZ perfect absorbers. 

In the next section we show that the magnitude of this MNZ absorption depends on the spin of the incident wave.

\section{Absorption efficiency is dependent on photonic-spin}

\begin{figure}
    \centering
    \includegraphics{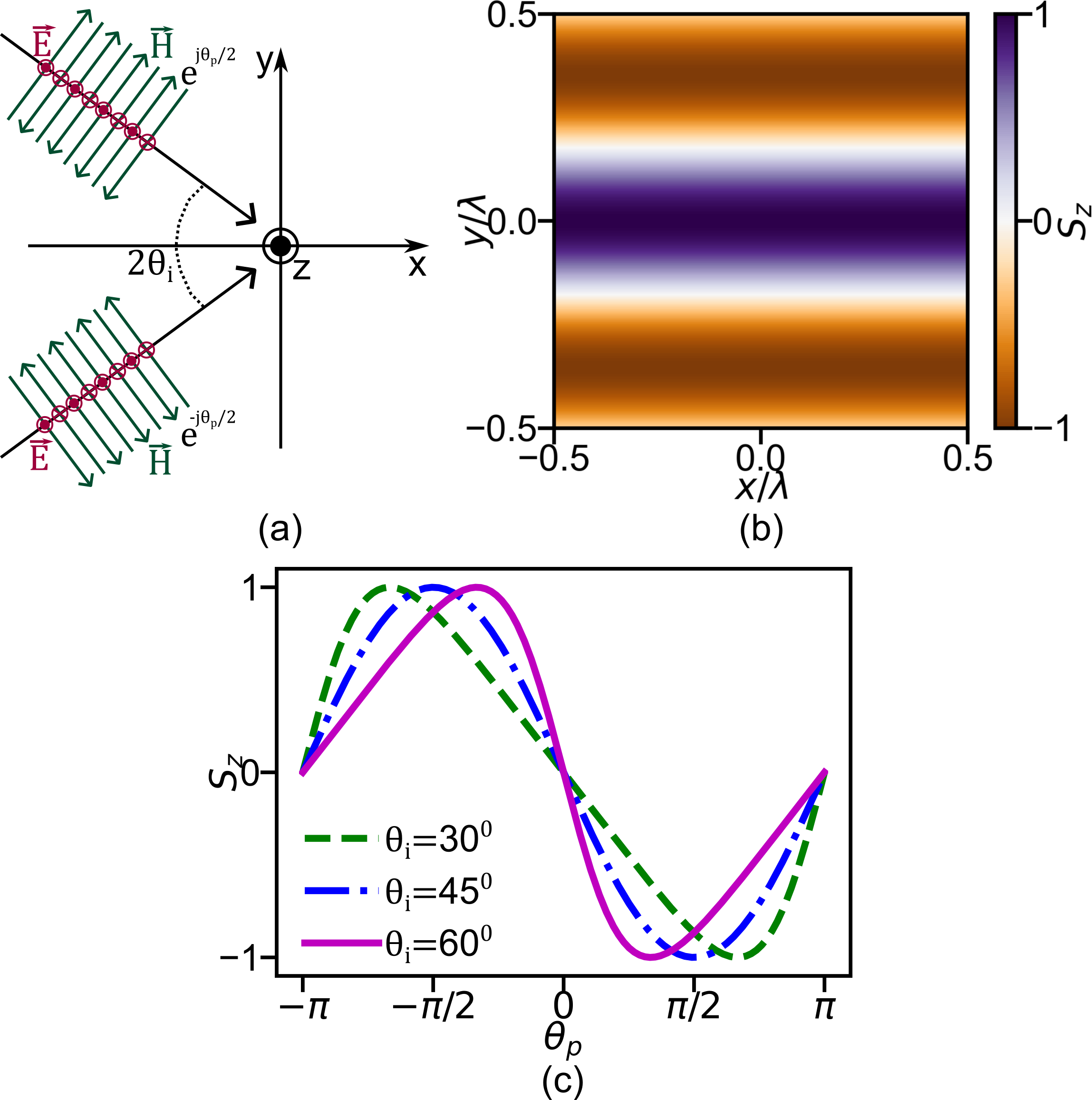}
    \caption{(a) Schematic representation of two obliquely incident plane waves (with $\theta_i$ as the angle of incidence with respect to $\hat{x}$-axis) to induce photonic spin at the origin. The incident plane waves have a phase difference of $\theta_p$. (b) Photonic spin profile within a region $-0.5\lambda\leq x,y\leq0.5\lambda$, where $\lambda$ is the free space wavelength at $f=15$ GHz. We consider $\theta_i=45^\circ$ and $\theta_p=-90^\circ$. (c) Photonic spin at the origin $x=0,y=0$ with respect to $\theta_p$ at different $\theta_i$ values.}
    \label{fig:fig2}
\end{figure}

In this section, we show that if the incident wave has a magnetic field spin, then its absorption by the gyromagnetic rod is dependent on its photonic-spin.

\subsection{Generating photonic-spin with obliquely incident plane waves}

Let us first see how a magnetic field spin can be generated by controlling the phase between two incident plane waves. Let two uniform plane waves be propagating in the $x-y$ plane with making an angle $+\theta_i$ and $-\theta_i$ with the $\hat{x}$ axis, as shown in Fig.~\ref{fig:fig2}(a). The equations for the magnetic fields of these two waves are given by,


\begin{multline}
    \vec{H}_{inc1}=H_0k_0(\sin(\theta_i)\hat{x}\\+\cos(\theta_i)\hat{y})e^\frac{j\theta_p}{2}e^{jk_0(\cos(\theta_i)x-\sin(\theta_i)y)}
\end{multline}
\begin{multline}
    \vec{H}_{inc2}=H_0k_0(-\sin(\theta_i)\hat{x}\\+\cos(\theta_i)\hat{y})e^\frac{-j\theta_p}{2}e^{jk_0(\cos(\theta_i)x+\sin(\theta_i)y)}
\end{multline}
$k_0$ is the free-space propagation constant and $\theta_p$ is the phase difference between the two incident waves. The resultant magnetic field is given by,
\begin{multline}
    \vec{H}_{inc} =2jH_0k_0\sin(\theta_i)\sin\left(\frac{\theta_p}{2}-k_0y\sin(\theta_i)\right)\hat{x}\\
    +2H_0k_0\cos(\theta_i)\cos\left(\frac{\theta_p}{2}-k_0y\sin(\theta_i)\right)\hat{y}
\end{multline}

It can be seen that the $\hat{x}$ and $\hat{y}$ components are out of phase. The magnetic field spin, $\vec{S}_h=S_z\hat{z}$, can be derived to be 

\begin{multline}
    S_z=\text{Im}(H_{inc,x}^*H_{inc,y}-H_{inc,x}H_{inc,y}^*)\\=-2H_0^2k_0^2\sin(2\theta_i)\sin(\theta_p-2k_0y\sin(\theta_i))
\end{multline}

By changing the phase difference $\theta_p$ it is possible to get a non-zero photonic spin at the center. Later, we will assume the gyromagnetic rod to be placed at the center. Figure~\ref{fig:fig2}(b) shows the spin profile over space for one such scenario with the angle of incidence $\theta_i=45^\circ$ and the phase difference $\theta_p=-90^\circ$. By changing $\theta_p$, the normalized spin at the center \textit{\textbf{($x=0,y=0$)}} can be varied between +1 and -1, as shown in Fig.~\ref{fig:fig2}(c). Thus, we can use two out-of-phase waves to generate spin in the incident wave and analyze its scattering and absorption properties. 

\subsection{Spin dependent absorption in MNZ gyromagnetic cylinder}

\begin{figure}
    \centering
    \includegraphics[width=\columnwidth]{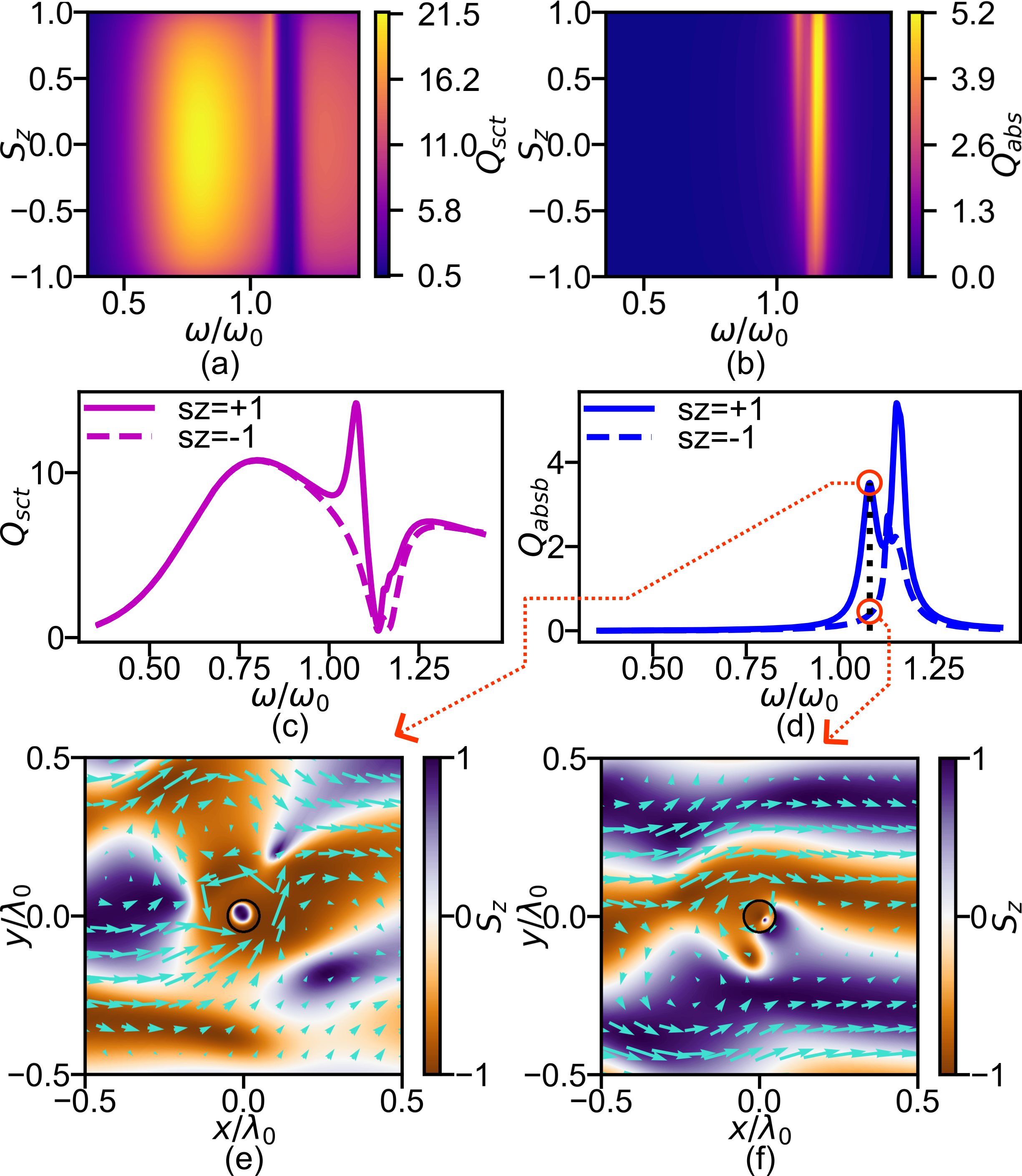}
    \caption{Colormap representation of efficiencies (a) $Q_{sct}$ and (b) $Q_{abs}$ as a function of frequency and spin. (c) $Q_{sct}$ and (d) (db) $Q_{abs}$ with respect to frequency variation at spin values $S_z=\pm1$. Color-representation of photonic spin and Poynting vector representation using arrows corresponding to (e) spin-match and (f) spin-mismatch conditions. Observations made at frequency $\omega=1.08\omega_0$, where spin-dependent absorption is maximum.}
    \label{fig:fig3}
\end{figure}

We will now show that due to the inherent spin of the mode supported inside the cylinder, its absorption efficiency becomes strongly dependent on the spin of the incident wave. For this, we consider a subwavelength $a=0.05\lambda_0$ gyromagnetic cylinder placed at the origin, with two incident waves as described in the previous sub-section falling on the cylinder. The spin of the incident wave is controlled by varying the phase difference between them. For computation of absorption efficiencies, the simple formula based on the summation of coefficients $A_m$s and $B_m$s cannot be used. We compute the absorption efficiencies by computing the net scattered and net incident power over a surface enclosing the rod. The incident Poynting vector is the vector sum of Poynting vector two incident waves. $\vec{P}_{inc} = (|\vec{P}_{inc1}|\cos(\theta_i)+|\vec{P}_{inc2}|\cos(\theta_i))\hat{x}$. The details of the computation are given in Appendix~\ref{ap:spin_add_effcns}. 

Figure~\ref{fig:fig3}(a) and (b) show the contour plots of scattering efficiencies and the absorption efficiencies, respectively, as a function of photonic spin and frequency. It can be seen that contour plots are asymmetric about $S_z=0$. The absorption efficiency $Q_{abs}$ peaks around $\omega=1.08\omega_0$, for $S_z=+1$, while diminishing to almost zero for $S_z=-1$. This indicates that the absorption strongly depends on the spin of the incident wave. Figure~\ref{fig:fig3}(c) and (d) shows the Poynting vector (arrows) of the incident plus the scattered wave, when the spin is $S_z=+1$ and $S_z=-1$, respectively. It can be seen that the Poynting vector swirls into the rod for matching spin of the incident wave and rod. Whereas, when the spin of the incident wave and the rod are of opposite sense, the Poynting vector goes around it in a more or less non-interacting way. This indicates that the interaction of the gyromagnetic cylinder with the incident waves is spin dependent.

\section{Non-reciprocal reflection due to spin-dependent absorption}

\begin{figure*}
    \centering
    \includegraphics{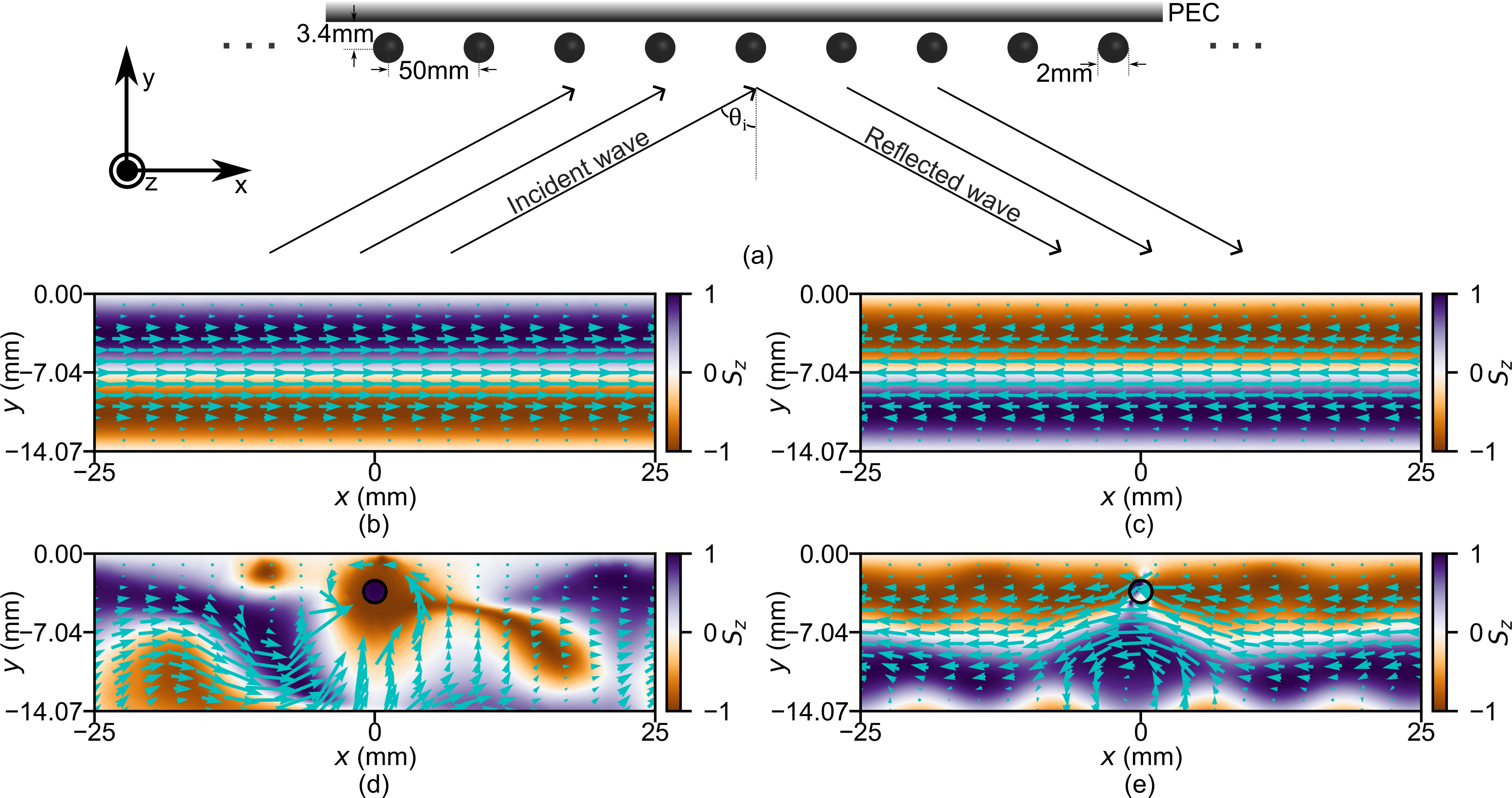}
    \caption{(a) Schematic of plane wave incidence and reflection from periodically arranged ferrite rods over PEC surface. The ferrite rods are embedded in the lossless RO4003C medium. Color-mapped photonic spin and Poynting vector representation using arrows corresponding (b) $\theta_i=+45^\circ$, no ferrite rod, (c) $\theta_i=-45^\circ$, no ferrite rod, (d) $\theta_i=+45^\circ$, with ferrite rod, and (e) $\theta_i=-45^\circ$, with ferrite rod. The material properties for the N32 ferrite used in this simulation are provided in Table~\ref{tab:n32_aprms}. The applied magnetic bias is $\vec{H}_0=1065\hat{z}$ Oe and the frequency of observation is $8$ GHz, which also corresponds to the frequency having maximum nonreciprocal absorption at $\theta_i=+45^\circ$ incidence.}
    \label{fig:fig4}
\end{figure*}

In the previous section we showed the spin-dependence of the subwavelength absorption. However, generating spin in the incident wave with superposition of coherent waves, with precise control over their phase difference, has limited practical applications. Fortunately, such a wave with transverse spin can be generated easily by superposition of incident TE wave and its reflection from a perfect electric conductor, as depicted in Figure~\ref{fig:fig4}(a). In this case, the spin profile depends on the angle of incidence. The direction of spin profile near the reflector can be reversed by swapping the direction of incident and reflected wave. This is shown by the spin profile near the reflector for $+45^\circ$ and $-45^\circ$ angle of incidence in Fig.~\ref{fig:fig4}(b) and (c). The Poynting vector flow in these cases is shown by the arrows. In Fig.~\ref{fig:fig4}(b), the net power flow is from left to right, while in Fig.~\ref{fig:fig4}(c) the direction of net power flow is reversed. These plots are without the rods.

\begin{figure}
    \centering
    \includegraphics{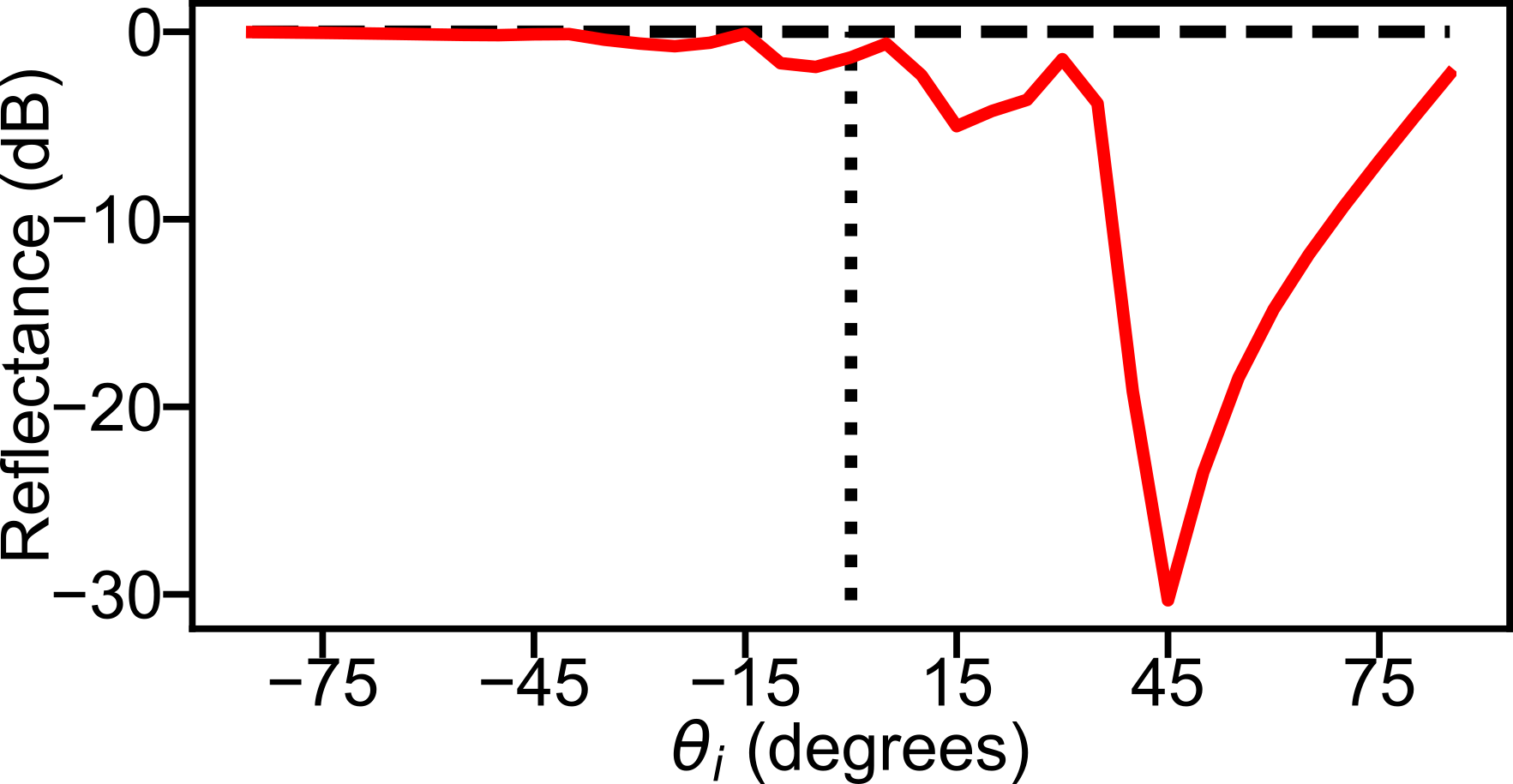}
    \caption{CST simulation result showing nonreciprocal photonic spin-dependent absorption for different angles of incidence at $8$ GHz and $\vec{H}_0=1065\hat{z}$ Oe.}
    \label{fig:NR_fer_refl}
\end{figure}

Now, if we place a series of rods at a position of maximum spin, we can see that for $+45^\circ$ of incidence, the Poynting vector just flows into the rod, indicating absorption. While for the reverse direction, the wave is not absorbed to that extent. This can be seen in Fig.~\ref{fig:fig4}(d) and (e). The resultant reflectance as a function of the angle of incidence is shown in Fig.~\ref{fig:NR_fer_refl}. We can clearly observe the breaking of reflection symmetry and non-reciprocal reflection. While we get high reflectance for negative values of $\theta_i$, at an incidence of $+45^\circ$, we get almost perfect absorption. This direction-dependent and non-reciprocal reflection can have several practical applications. 

\section{Waveguide isolator with spin-governed absorption}

\begin{table}
    \centering
    \caption{Exxelia Temex\textsuperscript{\textcopyright} N32 ferrite material parameters.}
    \begin{tabular}{cc}
    \hline
    \hline
        Parameter & Value\\
    \hline
        Saturation Magnetization ($4\pi M_s$) & $3161$ Gauss\\
        Land\'e g factor ($g$) & 2.3 \\
        Dielectric permittivity ($\epsilon_r$) & 13.3\\
        Linewidth ($\Delta H$) & $102$ Oe \\
        Loss tangent ($\tan\delta$) & $2.8\times10^{-4}$\\
    \hline
    \hline
    \end{tabular}
    \label{tab:n32_aprms}
\end{table}

A TE mode in a rectangular waveguide also supports a spin profile similar to that of the reflected TE wave which was discussed in the previous section. The spin profile for a dielectric waveguide was shown in \cite{Sarang_2018}, and for a metallic waveguide was reported in \cite{Sen_2022}. For the first-order mode in a waveguide, typically, the positive and negative spin is evenly distributed in upper and lower half of the cross-section (refer to Fig. 6 of \cite{Sen_2022}). The spin at the center is zero. At any offset point from the center of the axis of propagation, the sense of spin reverses with the reversal in the direction of propagation. If we place a subwavelength ferrite rod at an offset point, it should interact with an opposite sense of spins for the forward and the backward direction of propagation. Therefore if our claims of spin-dependent absorption are correct, it should absorb the wave only in one direction while allowing transmission in the opposite direction. 

\begin{figure}
    \centering
    \includegraphics[width=\columnwidth]{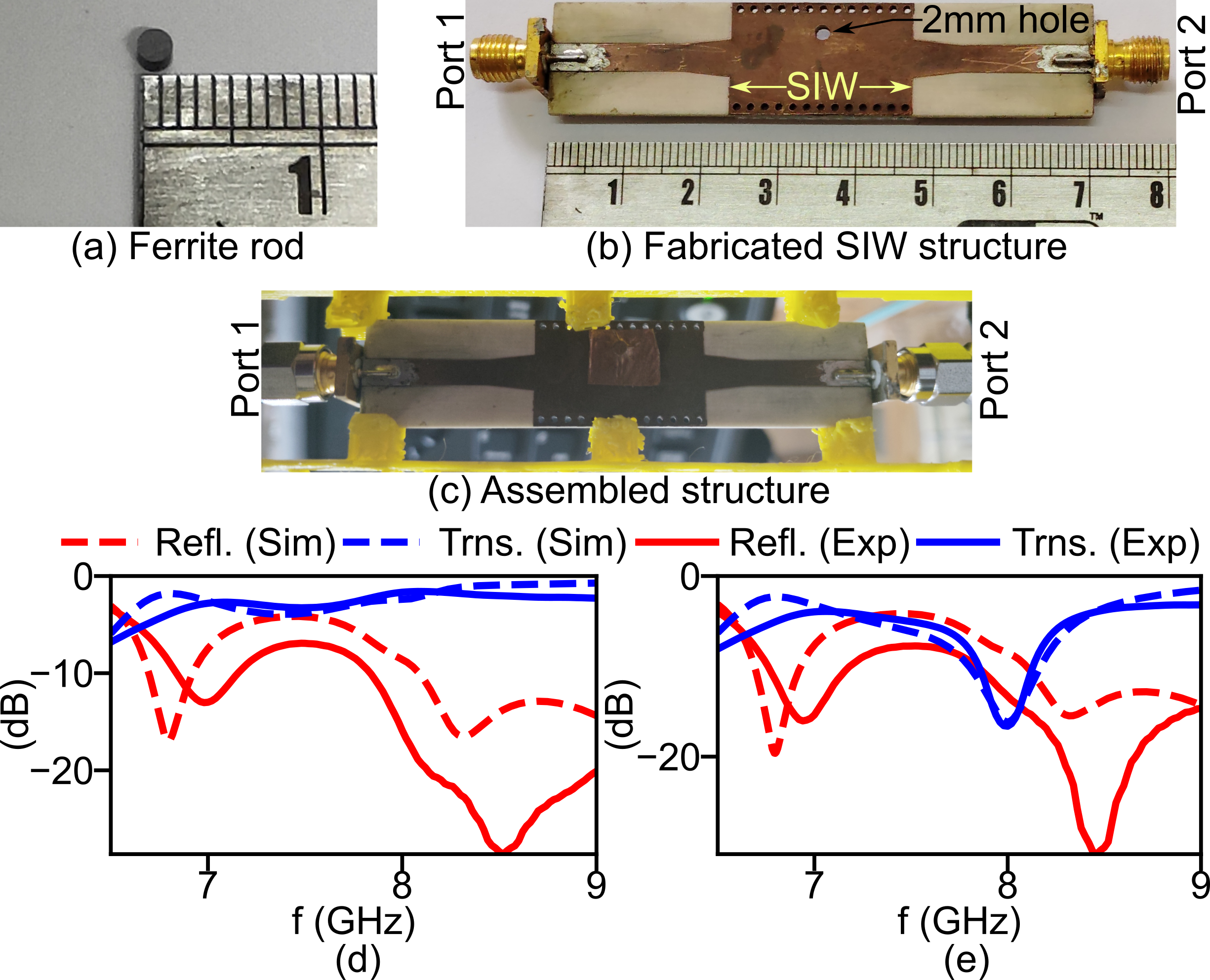}
    \caption{(a) N32 Ferrite rod of diameter $2$ mm. (b) Top-view of fabricated SIW structure with a $2$ mm hole to house the ferrite rod. (c) Photograph of the assembled structure. Transmission and reflection spectra corresponding to excitation at (d) Port-1 and (e) Port-2 and absorption at $8$ GHz.}
    \label{fig:fab_SIW_result}
\end{figure}

To validate our claims of spin-governed absorption, we fabricate a substrate-integrated waveguide (SIW) on Rogers\textsuperscript{\textcopyright} RO4003C substrate of thickness $1.524$ mm and dielectric constant $3.55$, and drill a small hole of $2$ mm diameter as shown in Fig~\ref{fig:fab_SIW_result}(b). We then insert a small N32 Spinel ferrite rod procured from Exxelia Temex\textsuperscript{\textcopyright}, of specifications mentioned in Table~\ref{tab:n32_aprms} and shown in Fig.~\ref{fig:fab_SIW_result}(a) into the hole and cover it with copper tape on both sides. The final SIW with a ferrite rod embedded in it is shown in Fig.~\ref{fig:fab_SIW_result}(c). We bias the ferrite rod with $\vec{H}=+910\hat{z}$ Oe. The measured results of transmission and reflection in the forward direction (port-1 being excited) are shown in Fig.~\ref{fig:fab_SIW_result}(d). It can be seen that it transmits in the forward direction. While for the backward direction, it blocks transmission at around 8 GHz, as shown in Fig.~\ref{fig:fab_SIW_result}(e). For this ferrite material, 8 GHz is near the MNZ frequency corresponding to the applied magnetic bias. It should also be noted that even when the transmission is low, the reflection does not increase, indicating absorption of the incident wave. By reversing the direction of bias, the direction of transmission and absorption can be switched. Moreover, by changing the strength of magnetic bias, the frequency of absorption can be tuned. Please refer to the additional files for a video of this demonstration. At $8$ GHz, the free space wavelength is $37.5$ mm, while in the bulk substrate Rogers\textsuperscript{\textcopyright} RO4003C of dielectric constant 3.55, its is $19.9$ mm. The guided wavelength in SIW is more than this. Therefore by using a small rod of 2 mm diameter we have demonstrated the subwavelength nature of this spin-governed absorption. This experimental verification also demonstrates the feasibility of using spin-dependent absorption for realizing compact substrate integrated isolators.

\section{Conclusion}

We have reported that a small subwavelength gyromagnetic rod exhibits spin-dependent absorption. We have analytically shown spin-dependent nature of scattering and absorption efficiencies of such rods. The reported claims are experimentally validated by demonstrating directional absorption in SIW waveguide. 

Even though the presented formulation is done for gyromagnetic materials, and the experimental demonstration is done at microwave frequencies, the reported phenomena will be valid for gyroelectric epsilon-near-zero materials at optical frequencies as well. The results presented here, therefore, introduces a novel approach for realizing compact, integrated, passive isolators at microwave as well as optical frequencies.

\begin{acknowledgments}
    We wish to acknowledge the financial support provided by the Prime Minister's Research Fellowship (PMRF), Govt. of India, which was critical for the experimental part of this work. We thank Mr. Jagannath Mukhi of the E\&ECE Department, IIT Kharagpur, for his invaluable technical support in fabricating and assembling the waveguide isolator prototype. 
\end{acknowledgments}

\appendix

\section{Scattering coefficients for gyromagnetic cylinder}
\label{ap:unknown_coefficients}

In accordance with our selected TE mode of wave propagation, the electric field component will be directed along the axis of the cylinder, that is the $\hat{z}$-axis. Using Jacobi-Anger expansion, we can write the electric field component corresponding to the wave incidence, scattering and absorption as
\begin{equation}
    E_z^i=E_0\sum_{m=-\infty}^\infty j^mJ_m(k_d\rho)e^{jm\phi}
\end{equation}
\begin{equation}
    E_z^s=E_0\sum_{m=-\infty}^\infty A_mj^mH_m^{(1)}(k_d\rho)e^{jm\phi}
\end{equation}
\begin{equation}
    E_z^t=E_0\sum_{m=-\infty}^\infty B_m j^mJ_m(k_f\rho)e^{jm\phi}
\end{equation}
Here $A_m$ and $B_m$ are the scattering and absorption coefficients, respectively. Superscripts, $i$, $s$ and $t$ correspond to wave incidence, scattering, and absorption, respectively.
As now we have the electric field equations for the three scenarios, we can find the corresponding magnetic field vector $\vec{H}$ using the Maxwell's equation $\nabla\times\vec{E}=-\partial(\mu_0\overset{\leftrightarrow}{\mu}_r\cdot\vec{H})/\partial t = j\omega \mu_0\overset{\leftrightarrow}{\mu}_r\cdot\vec{H}$ considering the basis function $\exp(j\vec{k}\cdot\vec{r}-j\omega t)$.  
\begin{equation}
    H_\rho^i = E_0\sum_{m=-\infty}^{\infty} \frac{m}{\omega\mu_0\rho} j^m J_m(k_d\rho) e^{jm\phi}
\end{equation}
\begin{equation}
    H_\phi^i = E_0\sum_{m=-\infty}^{\infty} \frac{jk_d}{\omega\mu_0} j^m J_m^\prime(k_d\rho) e^{jm\phi}
\end{equation}
\begin{equation}
    H_\rho^s = E_0\sum_{m=-\infty}^{\infty} \frac{m}{\omega\mu_0\rho} A_m j^m H_m^{(1)}(k_d\rho) e^{jm\phi}
\end{equation}
\begin{equation}
    H_\phi^i = E_0\sum_{m=-\infty}^{\infty} \frac{jk_d}{\omega\mu_0} A_m j^m H_m^{(1)^\prime}(k_d\rho) e^{jm\phi}
\end{equation}
\begin{equation}
    H_\rho^t = E_0\sum_{m=-\infty}^{\infty}  B_m j^m  \left(\frac{m\mu^\prime J_m(k_f\rho)-k_f\rho\kappa^\prime J_m^\prime(k_f\rho)}{\omega\rho\mu_0\mu_{eff}\mu^\prime}\right) e^{jm\phi}
\end{equation}
\begin{equation}
    H_\phi^t = E_0\sum_{m=-\infty}^{\infty}  j B_m j^m \left(\frac{-m\kappa^\prime J_m(k_f\rho)+k_f\rho\mu^\prime J_m^\prime(k_f\rho)}{\omega\rho\mu_0\mu_{eff}\mu^\prime}\right) e^{jm\phi}
\end{equation}
We will apply the boundary conditions $E^i_{z|\rho=a}+E^s_{z|\rho=a}=E^t_{z|\rho=a}$ and $H^i_{\phi|\rho=a}+H^s_{\phi|\rho=a}=H^t_{\phi|\rho=a}$ to obtain two linear equations. Solving the two equations will give us the coefficients $A_m$ and $B_m$ as
\begin{widetext}
\begin{equation}
\label{eq:ap_am}
    A_m=-\frac{m\kappa^\prime J_m(k_da)J_m(k_fa)+a\mu^\prime(\mu_{eff}k_dJ_m^\prime(k_da)J_m(k_fa)-k_fJ_m(k_da)J_m^\prime(k_fa))}{m\kappa^\prime H_m^{(1)}(k_da)J_m(k_fa)+a\mu^\prime(\mu_{eff}k_dH_m^{(1)\prime}(k_da)J_m(k_fa)-k_fH_m^{(1)}(k_da)J_m^\prime(k_fa))}
\end{equation}
\begin{equation}
    B_m=\frac{ak_d\mu_{eff}\mu^\prime(H_m^{(1)\prime}(k_da)J_m(k_da)-H_m^{(1)}(k_da)J_m^\prime(k_da)}{m\kappa^\prime H_m^{(1)}(k_da)J_m(k_fa)+a\mu^\prime(\mu_{eff}k_dH_m^{(1)\prime}(k_da)J_m(k_fa)-k_fH_m^{(1)}(k_da)J_m^\prime(k_fa))}
\end{equation}
\end{widetext}

As we now have the information about field components outside the ferrite rod (incident and scattered) in the region $\rho>a$, we can compute the scattered $W_{sct}$ and absorbed power $W_{abs}$ at a distance $\rho=a_c$ as
\begin{equation}
    W_{sct}=\int_0^{2\pi}\frac{1}{2}\text{Re}(E_z^{^*s}H_\phi^s)\rho d\phi_{|\rho=a_c}
\end{equation}
\begin{equation}
    W_{abs}=-\int_0^{2\pi}\frac{1}{2}\text{Re}((E_z^{s*}+E_z^{i*}) (H_\phi^s+H_\phi^i))\rho d\phi_{|\rho=a_c}
\end{equation}
Correspondingly, the absorption and scattering cross section can be computed using
\begin{equation}
    C_{sct}=\frac{W_{sct}}{|\vec{P}_{inc}|}
\end{equation}
\begin{equation}
    C_{abs}=\frac{W_{abs}}{|\vec{P}_{inc}|}
\end{equation}
Where $|\vec{P}_{inc}|$ can be simply expressed as $(E_0^2\cos(\theta_i)/\eta)$ and $\eta=\sqrt{\mu/\epsilon}$. Finally we can compute the scattering $Q_{sct}$ and absorption efficiencies $Q_{abs}$ using
\begin{equation}
\label{eq:q_sct}
    Q_{sct}=\frac{C_{sct}}{2a}
\end{equation}
\begin{equation}
\label{eq:q_absb}
    Q_{abs}=\frac{C_{abs}}{2a}
\end{equation}

\section{Computing the effect of spin in absorption efficiency}
\label{ap:spin_add_effcns}

Considering two obliquely incident plane waves incident toward the origin with angles $\pm\theta_i$ and phase difference $\theta_p$. The electric field vector for the waves can be written as
\begin{gather}
    \vec{E}_{inc1}=E_0\sum_{m=-\infty}^\infty j^mJ_m(k_d\rho)e^{j(m(\phi+\theta_i)+0.5\theta_p)}\hat{z}\\
    \vec{E}_{inc2}=E_0\sum_{m=-\infty}^\infty j^mJ_m(k_d\rho)e^{j(m(\phi-\theta_i)-0.5\theta_p)}\hat{z}
\end{gather}
Here, $E_0$ is a constant and will be considered to be $1$ V/m during computation for simplification. The corresponding magnetic field vectors can be written as
\begin{multline}
    \vec{H}_{inc1}=E_0\sum_{m=-\infty}^\infty \frac{mj^m}{\rho\omega\mu_0}J_m(k_d\rho)e^{j(m(\phi+\theta_i)+0.5\theta_p)}\hat{\rho}\\+\sum_{m=-\infty}^\infty\frac{j^{m+1}k_d}{\omega\mu_0}J_m^\prime(k_d\rho)e^{j(m(\phi+\theta_i)+0.5\theta_p)}\hat{\phi}
\end{multline}
\begin{multline}
    \vec{H}_{inc2}=E_0\sum_{m=-\infty}^\infty \frac{mj^m}{\rho\omega\mu_0}J_m(k_d\rho)e^{j(m(\phi-\theta_i)-0.5\theta_p)}\hat{\rho}\\+E_0\sum_{m=-\infty}^\infty\frac{j^{m+1}k_d}{\omega\mu_0}J_m^\prime(k_d\rho)e^{j(m(\phi-\theta_i)-0.5\theta_p)}\hat{\phi}
\end{multline}
Similarly, the scattered field components can be written as
\begin{gather}
    \vec{E}_{sct1}=E_0\sum_{m=-\infty}^\infty A_mj^mH_m^{(1)}(k_d\rho)e^{j(m(\phi+\theta_i)+0.5\theta_p)}\hat{z}\\
    \vec{E}_{sct2}=E_0\sum_{m=-\infty}^\infty A_mj^mH_m^{(1)}(k_d\rho)e^{j(m(\phi-\theta_i)-0.5\theta_p)}\hat{z}
\end{gather}
\begin{multline}
    \vec{H}_{sct1}=E_0\sum_{m=-\infty}^\infty A_m\frac{mj^m}{\rho\omega\mu_0}H_m^{(1)}(k_d\rho)e^{j(m(\phi+\theta_i)+0.5\theta_p)}\hat{\rho}\\+E_0\sum_{m=-\infty}^\infty A_m\frac{j^{m+1}k_d}{\omega\mu_0}H_m^{(1)\prime}(k_d\rho)e^{j(m(\phi+\theta_i)+0.5\theta_p)}\hat{\phi}
\end{multline}
\begin{multline}
    \vec{H}_{sct2}=E_0\sum_{m=-\infty}^\infty A_m\frac{mj^m}{\rho\omega\mu_0}H_m^{(1)}(k_d\rho)e^{j(m(\phi-\theta_i)-0.5\theta_p)}\hat{\rho}\\+E_0\sum_{m=-\infty}^\infty A_m\frac{j^{m+1}k_d}{\omega\mu_0}H_m^{(1)\prime}(k_d\rho)e^{j(m(\phi-\theta_i)-0.5\theta_p)}\hat{\phi}
\end{multline}
We can now find the combined incident and scattered fields as $\vec{E}_{inc}=\vec{E}_{inc1}+\vec{E}_{inc2}$, $\vec{H}_{inc}=\vec{H}_{inc1}+\vec{H}_{inc2}$, $\vec{E}_{sct}=\vec{E}_{sct1}+\vec{E}_{sct2}$, and $\vec{H}_{sct}=\vec{H}_{sct1}+\vec{H}_{sct2}$. The constant $A_m$ can be computed using eq.~\ref{eq:ap_am}, as it is independent of both $\theta_i$ and $\theta_p$. We can compute the corresponding scattering $Q_{sct}$ and absorption efficiencies $Q_{abs}$ using eq.~(\ref{eq:q_sct}) and (\ref{eq:q_absb}).

\bibliographystyle{ieeetr}
\renewcommand \bibname{References}
\bibliography{reference.bib}

\end{document}